\documentclass[%
 reprint,
 amsmath,amssymb,
 aps,
]{revtex4-2}
\usepackage{graphicx}
\usepackage{dcolumn}
\usepackage{bm}
\usepackage{xcolor}
\usepackage{soul}
\usepackage{makecell}
\usepackage{multirow}
\usepackage[compact]{titlesec}    

\begin{document}


\title{Spin wave behavior of a novel hopfion-like chiral state in Co/Pt nanodiscs}

\author{Nimisha Arora}
 \altaffiliation[Corresponding author:]{narora1.physics@gmail.com}
 \affiliation{%
Department of Physics, Indian Institute of Technology Delhi, 110016, India
}%
\author{Yogesh Kumar}%
 \altaffiliation[Corresponding author:]{yogeshmalya111@gmail.com}
\affiliation{%
Department of Physics, Indian Institute of Technology Delhi, 110016, India
}%
\author{Pintu Das}%
 \email{pintu@physics.iitd.ac.in}
\affiliation{%
Department of Physics, Indian Institute of Technology Delhi, 110016, India
}
\date{\today} 

\begin{abstract}
This work discusses the rich phase diagram of non-trivial chiral spin textures in confined ferromagnetic/heavy-metal (FM/HM) bilayer nanomagnets of circular cross-section. These spin textures are realized as a minimum-energy ground state during an external bias field sweep for a range of nanomagnet's diameter (d). Our study, based on micromagnetic simulations, has revealed a novel Hopfion-like state which can be stabilized for a wide range of diameters and external magnetic fields. We explored the dynamical characteristics of this novel Hopfion-like state under a transient magnetic field applied along the plane's perpendicular direction. Simulation results have demonstrated the excitation of nonreciprocal spin wave (SW) modes for this novel chiral state, in contrast to other stabilized chiral states. These modes are characterized as breathing and quantized radial modes, which also exhibit hybridization with azimuthal modes. The resonant SW modes have been used to demonstrate the switching from a Hopfion-like state to a skyrmion within a few nanoseconds of SW excitation. Furthermore, we establish a correlation between the behavior of excited SW modes as a function of external magnetic field strength and underlying chiral spin texture states.
\end{abstract}

\maketitle
\setlength{\parskip}{0pt}
\section{\label{sec:level1}Introduction}
In the realm of condensed-matter physics and materials science, the relentless pursuit of understanding and harnessing novel phenomena has guided researchers into the captivating world of spin textures~\cite{gobel2021beyond}. Spin textures arise due to the non-collinear distribution of the magnetization vector within the magnetic material.
Among these, chiral spin textures which prefer a particular handness, have emerged as distinctive entities which include domain wall~\cite{thiaville2012dynamics, ryu2013chiral}, radial and circular vortex~\cite{guslienko2008magnetic, antos2008magnetic}, skyrmion~\cite{liu2016skyrmions, finocchio2016magnetic, fert2017magnetic}, etc. in 2D, and hopfion~\cite{dzyloshinskiǐ1979localized, borisov2008stationary, kent2021creation} in 3D. They have garnered immense attention for their exceptional resilience to external perturbations owing to their topology (characterized by topological charge)~\cite{skyrme1961non}, scalability towards the nanoscale, and reconfigurability in response to external stimuli viz., magnetic field and electrical current~\cite{gobel2021beyond}. 
This remarkable property has catapulted chiral spin textures into the forefront of next-generation spintronic devices viz., innovative racetrack memories based on domain walls~\cite{parkin2008magnetic} or skyrmions~\cite{fert2013skyrmions, romming2013writing}, spin torque nano-oscillators~\cite{li2020spin, hu2022design, miriyala2019spin, liu2015dynamical}, magnetoresistive devices~\cite{pufall2007low, sluka2015spin}, logic devices~\cite{zhang2015magnetic, li2017magnetic}, and magnetic vortex oscillators~\cite{pribiag2007magnetic, mistral2008current}. Nevertheless, the seamless integration of these textures into devices requires controlled creation and manipulation of individual textures with an in-depth understanding of their dynamic (viz., spin wave) behavior~\cite{petti2022review, yu2021magnetic} under external perturbations.\par
 In this scientific pursuit, harnessing spin textures in the context of magnonics has recently ignited a new wave of theoretical and experimental efforts~\cite{wang2020effect, gypens2022commensurate, ding2015motion, song2019field, song2021spin, wang2020spin, vigo2022emergence}. These endeavors aim to explore the ability to stabilize distinct chiral spin textures and their interactions with SWs, offering the promise of novel device concepts and addressing longstanding challenges in the field of magnonics~\cite{yu2021magnetic, petti2022review}\. For instance, unique capability of spin texture in realizing the anisotropic excitation, spatial modulation of SWs and complex SW dispersion has allowed the developement of reprogrammable SW circuits~\cite{albisetti2018nanoscale}.
Furthermore, the dynamics of spin textures also play a pivotal role in generating 
 and propagating short-wavelength SWs~\cite{gypens2022commensurate, ding2015motion, song2021spin}. The tunability of the frequency and wavelength of these emitted SWs via excitation field presents a promising avenue for developing reprogrammable nanoscale SW sources.
As a matter of fact, aforementioned properties are highly dependent on the underlying spin configuration which emerge as the ground state of magnetic materials through a complex interplay of fundamental forces, including the symmetric Heisenberg exchange interaction, the asymmetric Dzyaloshinskii-Moriya (DM) interaction, and perpendicular anisotropy. This interplay results in a non-uniform energy landscape, and the second-order spatial derivative of this energy landscape dictates the specific configuration of these intriguing spin textures. In confined nanomagnets composed of ferromagnetic-heavy metal layered structures, additional factors such as shape, size, and the dipolar field emanating from the nanomagnet itself play a pivotal role in determining the resulting spin configurations~\cite{sampaio13, pollard2017observation, boulle2016room}. 
Thus, an in-depth understanding of the different energies is essential for achieving precise control and tunability in the nucleation, annihilation, and manipulation of these textures.\par
In general, dipolar interactions may play a significant role in stabilizing such spin textures in confined nanostructures.
While some studies~\cite{yu2012magnetic, castillo2019twisted, vigo2021spin, legrand2018,Lemesh17} have revealed that chiral textures such as Bloch skyrmions and twisted vortices can be stabilized by dipolar interactions alone, so far the role of dipolar interactions in presence of Heisenberg and anisotropic DM-exchange interactions in confined nanostructures remains to be properly understood.
Here we report systematic investigation of the role of dipolar interactions in stabilizing distinct chiral spin textures by varying the diameter of an ultrathin Co/Pt circular nanomagnet. Our study unravels a rich phase diagram of distinct chiral spin textures for a range of external magnetic field and nanomagnet's diameter (d). We report the presence of a novel spin texture whose spin configuration show significant discrepancies with the earlier reported chiral textures in 2D nanomagnet.
We extend the investigation to explore the SW behavior within this novel chiral spin texture, highlighting both similarities and distinctive features in comparison to reported chiral spin textures stabilized in ultrathin nanomagnets. Results based on our SW study demonstrate a sharp switching from this novel chiral state to a skyrmion state of opposite polarity induced via resonant excitation of the Eigen SW mode. The observed power-efficient magnetic state switching with opposite core polarity unveils promising avenues for advancing our understanding of these intriguing phenomena and their possible application in power-efficient memory devices and reconfigurable magnonics.


\section{\label{sec:level2}Micromagnetic Simulation}
For the investigations, we carried out detailed micromagnetic simulations for the individual circular-shaped nanomagnet, mimicking a Co/Pt bilayer system. For the simulations, geometrical parameters (thickness, \(t: 0.6\)\,nm and diameter range, \(d: 40\) to \(260\)\,nm) and material parameters (uniaxial anisotropy, \(K_{\rm{u}}: 0.52 \times 10^6\,\rm{J/m^3}\) and i-DMI, \(D_{\rm{int}}: 3\,\rm{mJ/m^2}\)) were taken from the experimentally reported values of the Co/Pt bilayer system in the literature \cite{sampaio13,TALAPATRA2018481}. These material parameters have earlier been shown to host the chiral spin textures for ultrathin Co/Pt bilayer nanomagnets. Here, the geometrical parameter (\(2R\)) is refined to realize the diverse non-trivial chiral magnetic states as a minimum energy ground state.\par
Accordingly, the minimum energy ground state is achieved by numerically solving the LLG equation using the fourth-order Runge-Kutta method in finite difference discretization-based (open-source) GPU-accelerated software M{\scriptsize U}M{\scriptsize AX}3~\cite{Vansteenkiste14, WinNT}. Here, the evolution of the space and time-dependent magnetization vector, \(\vec{m}(\vec{r}, t)\), is calculated at each cell of the discretized geometry by employing the Landau-Lifshitz-Gilbert equation (Eq.~\ref{eq: LLG}).
\begin{equation}\label{eq: LLG}
    \vec{\tau}_{\rm{LL}} = \frac{\partial\vec{m}}{\partial t} = \gamma_{\rm{LL}}\frac{1}{1+\alpha^2}[\vec{m}\times\vec{B}_{\rm{eff}} + \alpha(\vec{m}\times(\vec{m}\times\vec{B}_{\rm{eff}}))]
\end{equation}
with $\tau_{\text{LL}}$ is the Landau-Lifshitz torque, $\vec{m}(\vec{r}, t)$ is the reduced magnetization vector of unit length, $\gamma_{\rm{LL}}$ is the gyro-magnetic ratio (rad/Ts), $\alpha$ is the dimensionless damping parameter, and $\vec{B}_{\rm{eff}}$ is the effective magnetic field.\\ Here, $ \vec{B}_{\rm{eff}} = \vec{B}_{\rm{ext}}+\vec{B}_{\rm{demag}}+\vec{B}_{\rm{exch}}+\vec{B}_{\rm{anis}}+ \vec{B}_{\rm{dm}}+\vec{B}_{\rm{therm}}+\vec{B}_{\rm{exc}}(t)+...$\\
with $\vec{B}_{\rm{ext}}$, $\vec{B}_{\rm{demag}}$, $\vec{B}_{\rm{exch}}$, $\vec{B}_{\rm{anis}}$, $\vec{B}_{\rm{dm}}$, $\vec{B}_{\rm{therm}}$, and $\vec{B}_{\rm{exc}}(t)$ are the external bias field, demagnetization field, exchange field, anisotropy field, Dzyaloshinskii-Moriya (DM) exchange field, thermal field. and time-varying excitation (magnetic) field, respectively.\\
For the micromagnetic simulations, the geometry is discretized into cuboidal cells of 1\,nm length, which is smaller than the exchange length (\(\sim 8\)\,nm) of Co. The experimentally reported values of saturation magnetization (\(M_{\rm{sat}} = 1.1 \times 10^{6}\,\rm{A/m}\)), exchange stiffness constant (\(A_{\rm{ex}} = 1.6 \times 10^{-11}\,\rm{J/m}\)), and damping coefficient (\(\alpha = 0.5\)) for Co are used throughout the study~\cite{sampaio13, Nagaosa13, Xing20, TALAPATRA18}. To determine the SW behavior, a reduced damping coefficient of 0.008 is employed. This reduction enables a prolonged precession of weak modes, thereby allowing for a detailed analysis of such modes.\par
Dynamical characteristics (SW modes) are investigated for a nanomagnet of diameter: \(180\)\,nm as a model system. The diameter of \(180\)\,nm is chosen as it entails the chiral spin configurations of our interest, namely, the novel Hopfion-like state. SWs are excited by perturbing the equilibrium magnetic state stabilized at a particular magnetic field (\(B_{\rm{ext, z}}\)) via an out-of-plane (OOP) temporal magnetic field pulse (see details in supplementary information). After the incidence of the pulse, the time evolution of the reduced magnetization vector (\(\vec{m}(t)\)) is recorded for \(10\)\,ns at a time-step (\(\Delta t\)) of \(5\)\,ps. To extract the frequencies of the excited SW modes, a fast Fourier transformation (FFT) is performed on the recorded \(\vec{m}(t)\) data. Thus, power spectra for SW modes are determined for a given magnetic state. The nature and the origin of the SW modes are analyzed by investigating the power and phase profiles of the excited SW modes at particular frequencies using a MATLAB code~\cite{arora2022excitation}.

\begin{figure}
\includegraphics[width=\linewidth]{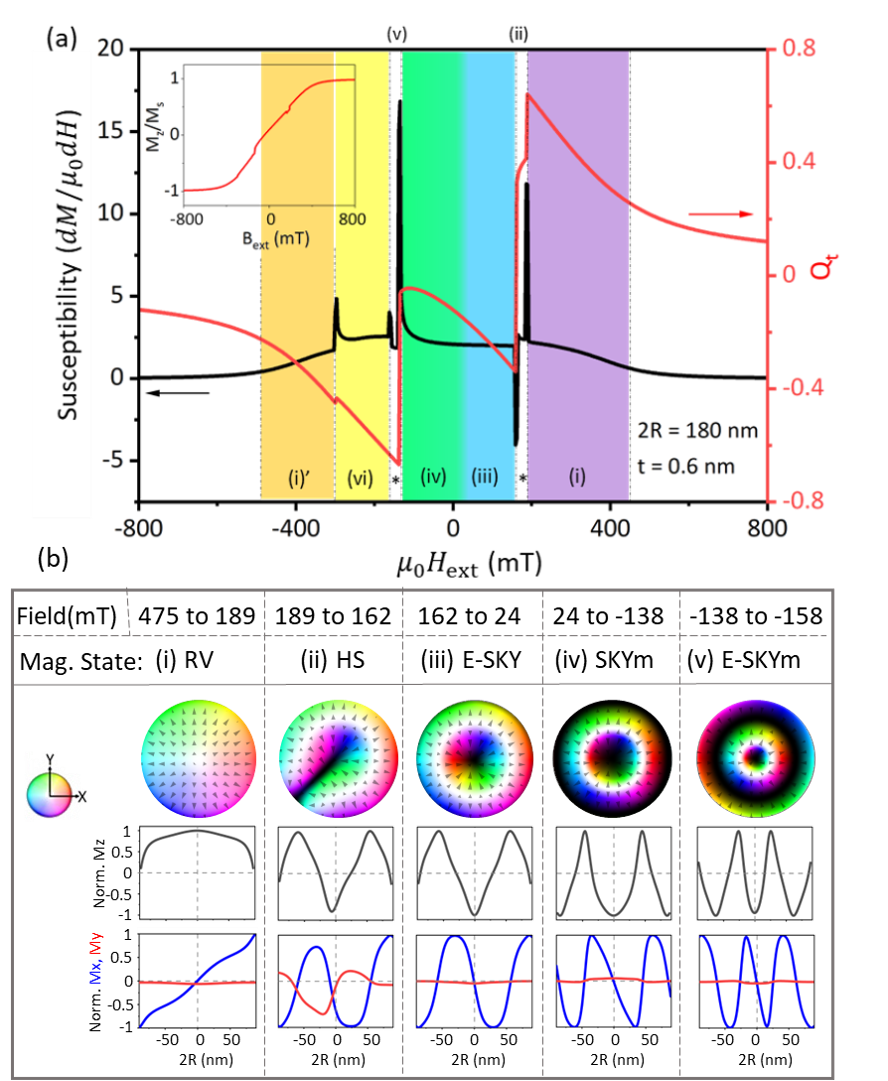}
\caption{\label{fig:1} (a) Response of susceptibility ($dM_{\rm{z}}/dB_{\rm{ext}}$) and topological charge during the downsweep of external magnetic field for Co/Pt circular nanomagnet of diameter (d): 180\,nm. Inset on the top left corner represents the corresponding response of normalized magnetization component ($M_{\rm{z}}/M_{\rm{s}}$) along the z-axis during the field sweep. 
(b) representative 2D magnetization profile of the observed magnetic states within their stability regime [highlighted in (a)]. Here, white and black color denotes the spins pointing in the surface normal direction, viz., +z and -z-axis, respectively; however, other colors represent magnetization in the nanomagnet plane. Arrows in the 2D magnetization image denote the spin orientation in the nanodisk plane.  Respective line profiles of the reduce magnetization components ($m_{\rm{x}}, m_{\rm{y}},$ and $m_{\rm{z}}$)cacross the geometrical center of the nanomagnet are shown below it's each respective 2D magnetization image in cartesian coordinate.}
\end{figure}

\section{\label{sec:level3}Results and Discussion}
To comprehend the influence of dipolar interaction on the energy-minimized ground state in Co/Pt circular nanomagnet, we conducted a quasi-static magnetic field dependent study for a diameter range of 40 to 300\,nm. The external magnetic field ($\mu_{\rm{0}}H_{\rm{ext}}$)  was applied along the z-axis and gradually downswept from 2 to -2\,T, in steps of 3\,mT. Figure\,\ref{fig:1}(a) shows representative response of susceptibility ($\chi$ or $dM_{\rm{z}}/\mu_0dH_{\rm{ext}}$) and topological charge ($Q_{t}$) corresponding to the underlying magnetic state for the representative case of diameter (d) = 180\,nm. The topological charge shown in Fig.\,\ref{fig:1}(a), is calculated using the following equation:
\begin{equation}\label{eq: top charge}
   Q_t= \frac{1}{4\pi}\int{d^d(\partial_x \vec{M} \times \partial_y \vec{M}).\vec{M}}
\end{equation}
where $\vec{M}$ is the unit vector of the magnetization.\\
Starting from the positive saturation, as the field is decreased below 475\,mT, the spins positioned at the boundary start to tilt towards the periphery. This tilt, forming an angle $\theta$ with the z-axis, is due to the interplay of competing DM, symmetric exchange, anisotropy, dipolar, and Zeeman interactions.
For 475\,mT\,$\leq \mu_0H_{\rm{ext}} \leq 189$\,mT, the total energy of the system is minimized by continuous increase of $\theta$. As a result, $M_{\rm{z}}/M_{\rm{s}}$ reduces almost linearly which is reflected as a observable slight gradient in $\chi$ (rwgion-i in Fig.\,\ref{fig:1}(a)). Simulation results suggest that in this field range, gradual tilting of spins leads to a smooth evolution into a radial vortex (RV) state with core magnetization parallel to $\mu_{\rm{0}}H_{\rm{ext}}$. During this evolution, associated topological charge $Q_{\rm{t}}$ smoothly varies from $\sim$\,0.25 to 0.7\,. The observed variation in $Q_{\rm{t}}$ can be accounted using the following expression~\cite{Zheng17}: 
\begin{equation}\label{eq: Qt due to rotation}
    Q_{\rm{t}} = p(1-cos\psi)/2
\end{equation}
Here, `p' denote the polarity of the spins at the center and $\psi$ denote the rotation of spins from the center to the boundary e.g., $\psi = \pi$ for skyrmion. The aforementioned expression is valid for vortices as well as k$\pi$\,skyrmions where additional rotation of spins at the edges is present due to the boundary imposed non-trivial DMI condition in such confined nanomagnets. The spin configuration of RV state along with the line profile of in-plane (IP) and OOP components of reduced magnetization across the geometrical center are shown in Fig.\,\ref{fig:1}(b.(i)). Such RV state is frequently observed in confined nanomagnets with i-DMI~\cite{siracusano2016magnetic}.\par

With further decrease of $\mu_{\rm{0}}H_{\rm{ext}}$, a few features as peak and dip are observed in $\chi$-$\mu_0H$ at field strengths of 189, 162, -135, -158, \& -296\,mT, respectively. Noticeable concurrent changes are also observed in $Q_{\rm{t}}$-$\mu_0H$ at the corresponding fields (see Fig.\,\ref{fig:1}).
A careful analysis of the local magnetization behavior reveals that the peak observed at 189\,mT corresponds to the transformation of underlying RV state to a precursor horse-shoe (HS)-like state~\cite{karakas18} which finally stabilizes to an extended-skyrmion (E-SKY)~\cite{Zheng17} state. 
E-SKY state stabilized in region-iii has $Q_{\rm{t}} <$\,1 ($Q_{\rm{t}}$ of SKY in thin film) which is due to the additional spin rotation at the edges and can be accounted from eqn\,\ref{eq: Qt due to rotation}.
As $\mu_{\rm{0}}H{\rm{ext}}$ is gradually reduced to zero, a constant $\chi$ response corresponding to linearly varying magnetization is observed with a concurrent decrease in the magnitude of $Q_{\rm{t}}$.  This behavior suggests a smooth evolution of the underlying magnetic state. However, the underlying spin texture is found to smoothly evolve into an another chiral state having three OOP polarized regions separated by two N\'{e}el-type domain walls of opposite chirality. This chiral state is well known as skyrmionium (SKYm) or 2$\pi$-skyrmion state~\cite{vigo2022emergence}. 
The 2D spin configuration of this state together with the corresponding line profile of the magnetization components are shown in Fig.\,\ref{fig:1}(b.(iv)). 
Remarkably, with further change of field towards the negative saturation, we observe the SKYm state transforming into a non-trivial, novel chiral state through the formation of another precursor state viz., the extended skyrmionium (E-SKYm) state. 
As shown in the Fig.\,\ref{fig:1}a, the formation of E-SKYm state is represented by a sharp peak in $\chi$ as well as a step in $Q_{\rm{t}}$ at -138\,mT. The 2D spin configuration and the line profiles of magnetization components are shown in Fig.\,\ref{fig:1}(b.(v)).\par
\begin{figure}
\includegraphics[width = \linewidth]{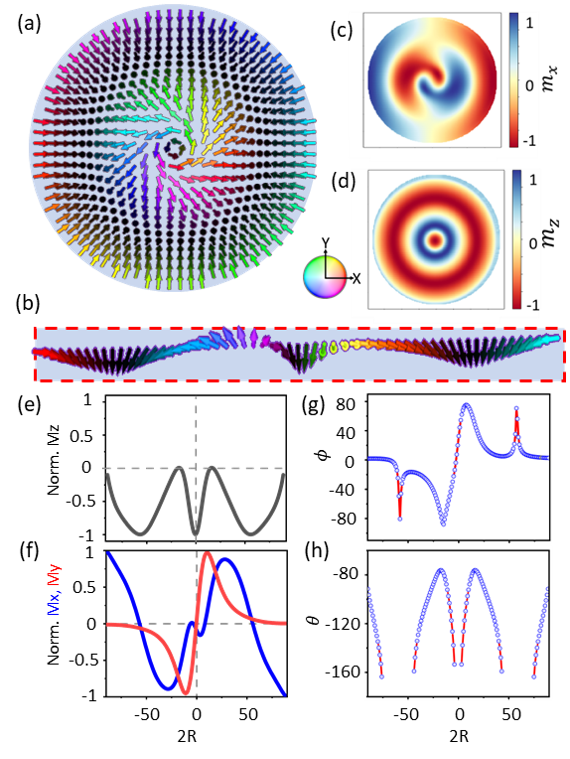}
\caption{\label{fig:2} (a) Magnetization profile of the novel hopfion-like state as 2D quiver plot. Color of the arrow is associated with different spin orientation in the plane of the nanomagnet as opted from the color wheel shown in bottom right corner.Whereas black and white color denote the spin orientation along inverse-surface normal and surface normal direction, respectively. (b) Cross-section view of the spin configuration across the geometrical center of the nanomagnet (radial direction, i.e., x-axis). (c, d) Spatial distribution of the reduced IP and OOP magnetization component of the hopfion-like state. Colorbars on the right denote the magnitude of the reduced magnetization components. (e, f) Line profile of the OOP and IP magnetization component across the geometrical center as extracted from (a) and (g, h) Line profile of the azimuthal and polar angle of the magnetization vector across the geometrical i.e., x-axis as calculated from (e) and (f)}
\end{figure}
The novel magnetic state stabilized within -158 to -296\,mT is illustrated in Fig.\,\ref{fig:2}(a, b) as a 2D quiver plot along with its cross-section view across the geometrical center. The spin configuration demonstrates significant deviation from the above studied states. 
Figure\,\ref{fig:2} also illustrates the 2D spatial distribution of the IP and OOP magnetization components of this magnetic state, along with the line profile of its reduced magnetization components ($M_{\rm{x, y, z}}/M_{\rm{s}}$), polar ($\theta$) and azimuthal ($\phi$) angles of the magnetization vector across the geometrical center. This particular magnetic state differs from the previously discussed states in the following ways:
1) The OOP magnetized regions at the nanomagnet center and near the edges are aligned parallel to each other, i.e., along the $\mu_0H_{\rm{ext}}$ (-z axis).
2) The domain wall separating these OOP magnetized regions exhibits a spiral rotation, as observed in Fig.\,\ref{fig:2}(a, b). This suggests the presence of a hybrid or twisted domain wall, displaying characteristics of mixed N\'{e}el and Bloch-type rotation.
The line profile analysis of the IP and OOP magnetization components across the geometrical center unequivocally confirms the presence of nonzero $M_x(r)$ and $M_y(r)$ components i.e., radial and transverse magnetization components, along the outward radial direction. This characteristic behavior is indicative of an additional degree of freedom, referred to as helicity ($\gamma$). Helicity measures the angle ($\phi$) subtended by the IP magnetization component on the radial axis, which is $\pm\pi/2$ in the case of a Bloch domain wall and 0 or $\pi$ in the case of a N\'{e}el domain wall. Interestingly, we observe a smooth variation in $\gamma$ from $\pi/2$ to 0 along the radial direction, indicative of hybrid domian wall rotation. Typically, i-DM interaction tends to favor N\'{e}el-type spin rotation, while bulk-DM interaction and dipolar interaction prefer Bloch-type spin rotation~\cite{yu2012magnetic, castillo2019twisted, vigo2021spin}. This suggests that the observed hybrid domain wall with variable helicity ($\gamma$) is a consequence of the interplay between these competing non-collinear interactions (i-DM and dipolar), particularly in the presence of strong perpendicular magnetic anisotropy. Similar hybrid spin rotation has been previously observed in skyrmions~\cite{Castillo19} and vortices~\cite{Verba20}, where the inclusion of long-range dipolar interactions in the energy function was found to play a crucial role.
It's worth noting that the chiral magnetic state shown in Fig.\,\ref{fig:2}(a) cannot be considered as a skyrmion or vortex-like state due to the stark difference in the relative alignment of its OOP magnetized regions. Interstingly, the spatial distribution of IP and OOP magnetization components in this novel magnetic state bears a striking resemblance to the magnetization components of a three-dimensional magnetic soliton known as a hopfion~\cite{kent2021creation}. To the best of our knowledge, there have been no reports of this type of magnetic state in FM/NM bilayer system. 
Given the significant similarity of its magnetization configuration to that of hopfion, we hereby label thisstate as the ``hopfion-like" state. This magnetic state proves to be robust over a wide range of magnetic field (-158 to -296\,mT, region-vi). As the magnetic field strength increases beyond -296\,mT, the hopfion-like state transfigures into a RV state (in region-i$^\prime$), with its core magnetization parallel to $\mu_0H_{\rm{ext}}$(-z-axis). This state is similar to the previously discussed RV in region-i. With further increases in the field strength along negative saturation, the spins begin to reorient along the magnetization field direction due to strong Zeeman interaction, eventually resulting in saturation beyond -473\,mT. 
\par
\begin{figure}
\includegraphics[width = \linewidth]{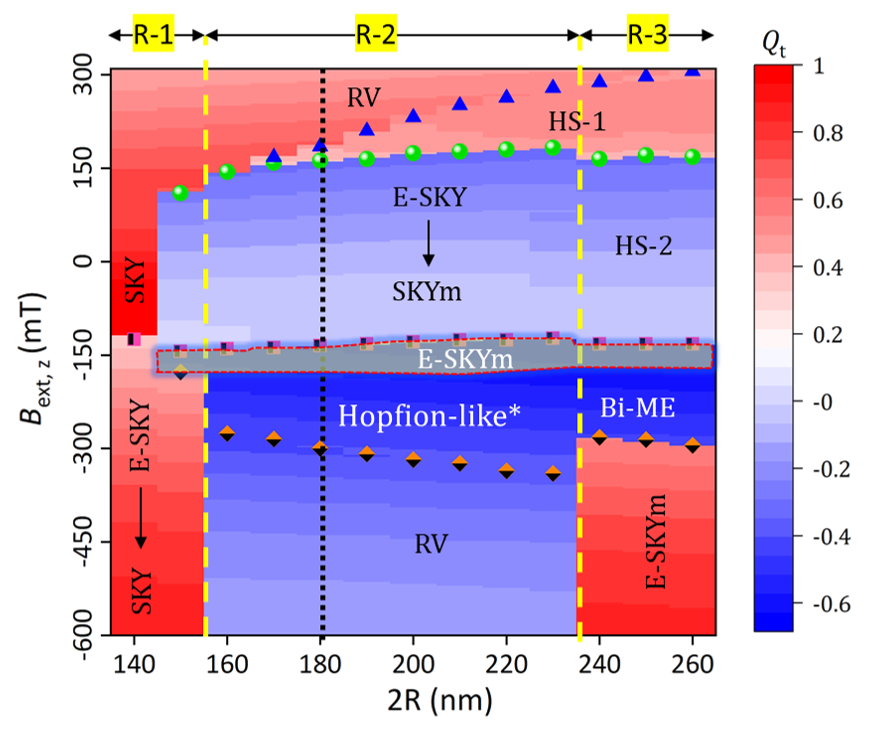}
\caption{\label{fig:3} Phase diagram of the stabilized magnetic state in Co/Pt circular nanomagnet for d: 140 to 260\,nm and magneteic field ($\vec{B}_{\rm{ext,z}}$): 300 to -600\,mT. Color in the phase diagram denote the associated topological charge with the underlying magnetic state. Colorbar on the right quantify the topological charge.}
\end{figure}
Furthermore, we have conducted the similar analysis for 2R ranging from 40 to 300\,nm in Co/Pt ultrathin nanomagnet. 
Figure\,\ref{fig:3} presents a phase diagram illustrating the observed magnetic phases for 140$\leq 2R\leq$260\,nm and 300$\leq \mu_0H_{\rm{ext}}\leq$-600\,mT. For $d<140$\,nm, only SKY and RV state is stabilized and therefore this diameter range is omitted in the phase diagram. The magnetic field range shown in Fig.\,\ref{fig:3} covers the full range of chiral magnetic states observed during magnetic field sweep within $\pm$\,2\,T, at steps of 3\,mT. The color-coding in the phase diagram represents the quantified $Q_t$ associated with the energy-minimized ground states. Sharp transitions in the color of the phase diagram i.e.,red to blue or vice versa, signify abrupt transitions between magnetic states. In contrast, gradual transitions are indicated by a continuous color gradient. Notably, while these transitions are evident in the internal spin configurations, they do not reflect as prominently in the associated $Q_t$, $\vec{M}$, and $\chi$ in all transitions e.g., for d=180\,nm. As a result, the phase diagram plotted using $Q_t$ does not provide clear demarcations for few magnetic phase transitions viz., E-sky to skym (d: 160 to 230\,nm) and E-sky to sky (d: 140 to 150\,nm). To highlight the observedtransitions, colorful markers have been overlaid on the phase diagram, segmenting regions of distinct magnetic states. These markers denote the position of peaks in the total energy derivative which correspond to the transitions between different magnetic states (see supplementary Figure\,SF1). 
For clarity, an additional phase diagram employing distinct colors to differentiate various magnetic states is shown in supplementary Fig.\,SF2. The formulation of this phase diagram is based on visual inspection of magnetic states during the magnetic field sweep.
Evidently from Fig.\,\ref{fig:3}, hopfion-like state is observed within the diameter (d) range of 155 to 235\,nm. Additionally, other chiral magnetic phases discussed for 2R = 180\,nm, are also present in this region. This region is labeled as R-2 in Fig.\,\ref{fig:3}, where moderate contribution of dipolar energy is present. Accordingly, phase diagram is segmented into three regions as follows:\\
1) Region-1 (R-1): 2R\,$<$ 155\,nm (with a nominal contribution of dipolar energy)\\
2) Region-2 (R-2): 155\,nm\,$<$ 2R\,$<$\,235 nm (with a moderate contribution of dipolar energy)\\
3) Region-3 (R-3): 2R\,$\geq$\,235\,nm (with a significant contribution of dipolar energy)\\
In R-1, all the chiral magnetic states observed in R-2, are present except HS and Hopfion-like state. However, in R-3, new magnetic states viz., type-2 horseshoe (HS-2) and Bi-meron-like (Bi-ME) are observed instead of E-SKY, SKYm and Hopfion-like state for 150\,mT$\leq\mu_0H_{\rm{ext}}\leq$-300\,mT, respectively. The 2D magnetization profile of these new magnetic states are shown in supplementary Fig.\,SF2 
\par
\begin{figure}
    \centering
    \includegraphics[width = \linewidth]{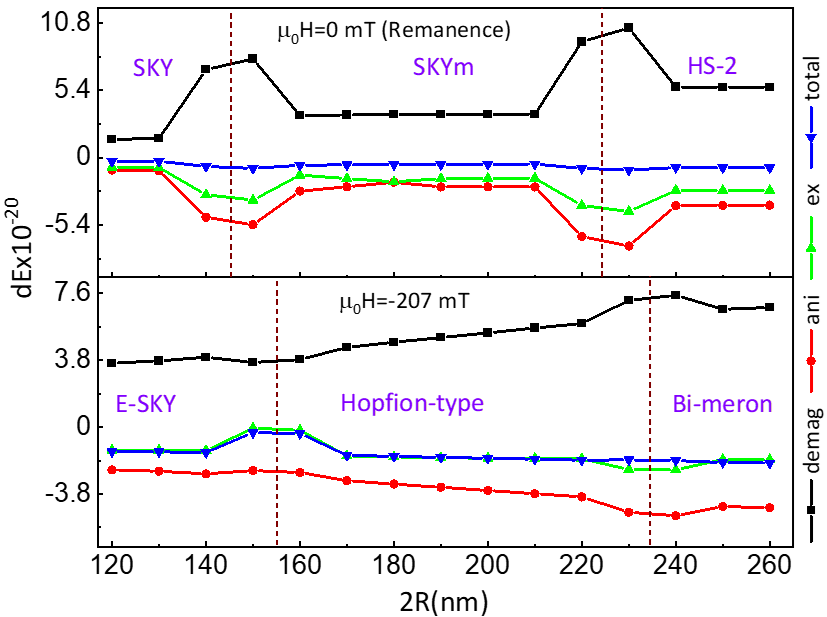}
    \caption{First order derivative of competing energies w.r.t diameter of nanomagnet at remanence (top) for external magnetic field strength ($\vec{B}_{\rm{ext, z}}$) of -207\,mT (bottom)}
    \label{fig:4}
\end{figure}
The observation of these different magnetic states for variable 2R is due to the variation in competing energies. To investigate the underlying energetic behind the stabilization of novel hopfion-like state, we have investigated the derivative of competing energies (dE) as a function of d. Figure\,\ref{fig:4} shows the representative response of dE vs. 2R at remanence and $\mu_0H_{\rm{ext}}$=\,-207\,mT, where novel hopfion-like state is absent and present, respectively. The corresponding E vs. 2R plot is shown in supp. Fig.\,SF2. Features like peaks and dips in the dE-d plot correspond to the increase and decrease in respective energies during transitions. Typically, the equilibrium magnetization configuration is majorly dominated by the particular energy which is minimized during the transition.  
In remanence, all competing energies except dipolar (demag) energy are minimized during the transitions which appear as dips in the Fig.\,\ref{fig:4}. Similar observation is made during the second transition at $\mu_0H_{\rm{ext}}$=-207\,mT.
Conversely, transition from E-Skym to hopfion-like state occur via minimization in demagnetization (dipolar) energy (\(E_{\text{demag}}\)) which appear as a dip in dE vs. 2R plot. We observe a simultaneous increase in total exchange (Heisenberg $+$ DM) and anisotropy energy. The increase in exchange energy may be attributed to the appearance of opposite-facing IP domains (inward and outward N\'{e}el domain walls) resulting from the sudden loss of the OOP magnetized (\(+m_{\text{z}}\)) region, earlier present in E-SKYm (refer to supplementary movie 1). This increase in exchange energy is compensated by the observed twisted spin rotation in the domain wall region. This twisting is indeed reflected as a decrease in dipolar energy, particularly for hopfion-like state. Further, the increase in anisotropy energy can be attributed to the increased domain wall width in the hopfion-like state. Thus, dipolar energy appears to play a crucial role in stabilizing a hopfion-like state where presence of hybrid-domain wall is supported by the minimization of dipolar (demag) energy. 
\par
\begin{figure*}[hbt!]
\includegraphics[width = \textwidth]{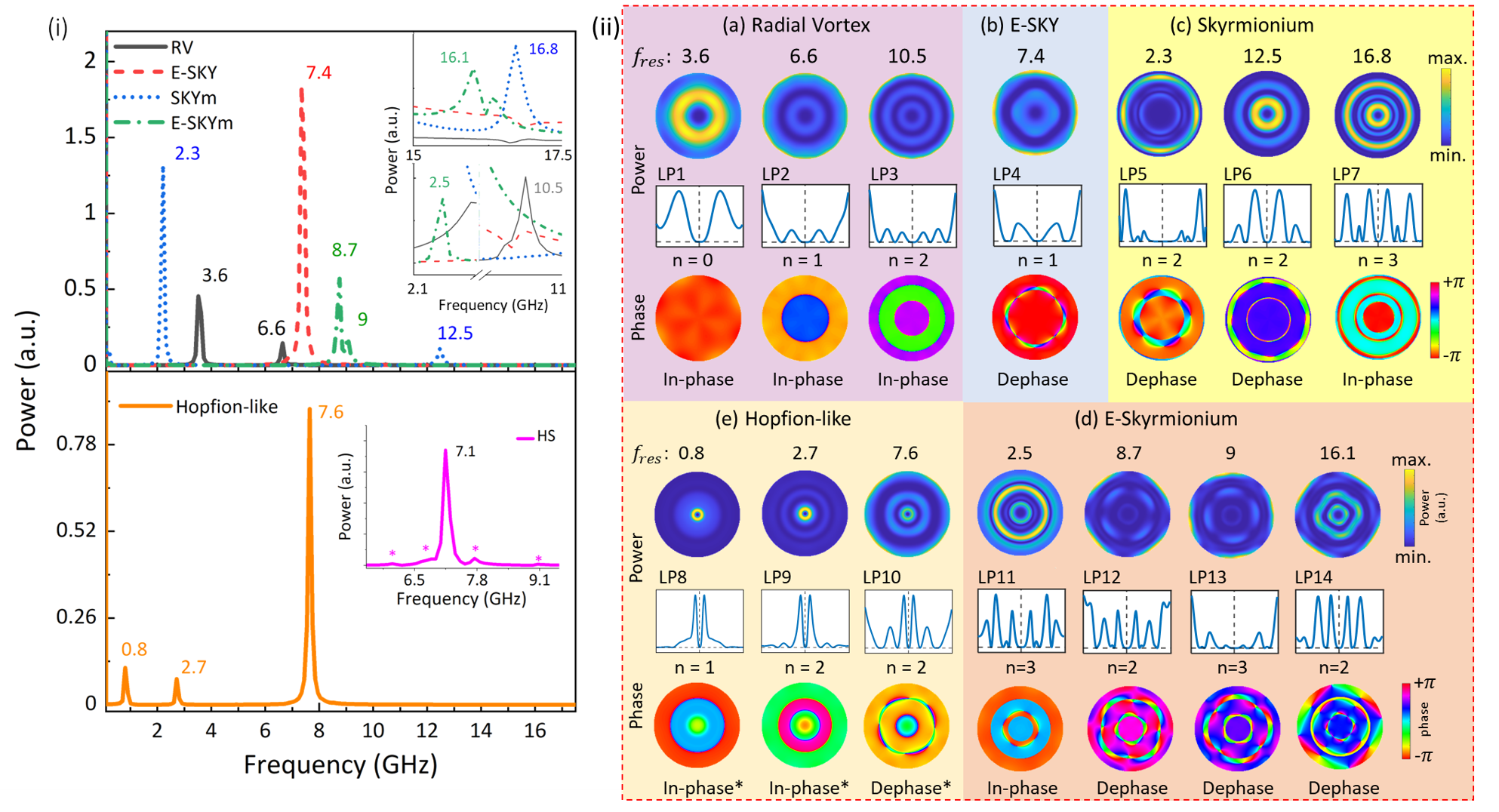}
\caption{\label{fig:5} (i) SW power spectra of the distinct non-trivial magnetic states at intermediate field strength within their respective field regime: top curve exhibit spectra for radial vortex (290\,mT), E-skyrmion (130\,mT), skyrmionium (-135\,mT) \& E-skyrmionium (-140\,mT), and bottom curve exhibit spectrum for hopfion-like state (-275\,mT). Insets in top right corner show the zoomed image of the spectra for frequency range:15 to 17.5\,GHz (top) and 10 to 11\,GHz (bottom) to highlight the presence of weaker SW modes in the studied magnetic states. Inset in bottom right corner shows the spectrum for metastable horshoe state (170\,mT). (ii) 2D spatial distribution of power and phase of the SW modes [shown in (i)] along with its respective line power profile across the geometrical center below each spatial power distribution image of (a) Radial vortex (290\,mT), (b) E-skyrmion (130\,mT), (c) Skyrmionium (-135\,mT), (d) E-skyrmionium (-140\,mT), and (d) hopfion-like (-275\,mT) state. Colorbar is shown at the bottom right of the image. Power is calculated from the FFT of the $\delta m_{\rm{z}}(t)$ fluctuations and imaged in arbitrary units for the best visualization (see details in supplementary info).}
\end{figure*}

The dynamic behavior of such chiral structures is of interest due to their unique capability in realizing anisotropic excitation~\cite{lan2015spin}, short-wavelength SWs~\cite{ding2015motion}, and topological magnon~\cite{wang2018topological} etc. In order to investigate the dynamic behavior in such chiral magnetic states, detailed analysis of SW behavior for the observed chiral states are carried out. 
For our analysis, an external transient magnetic field pulse was applied along along OOP, i.e., z- direction (see sec. II for details). This results in an excitation of SWs in GHz ferquencies carrying distinct characteristics for the underlying magnetic states. Figure\,\ref{fig:5}(i) shows the representative spectra calculated for the different magnetic states observed in the Co/Pt bilayer nanomagnet which we discussed above. The spectra were obtained by performing a fast Fourier transform on the time response of the spatially averaged magnetization component($m_{\rm{z}}$). Here we underline the important differences in the SW behavior for the novel Hopfion-like state which we observe for a wide range of $\mu_0H_{\rm{ext}}$ and $d$ as shown in $\mu_0H_{\rm{ext}}$-$d$ phase diagram in Fig.\,\ref{fig:3}.

Table-\ref{tab:my_label} outlines the distinct modes excited in different chiral magnetic states. The three excited modes observed for the RV state are due to the fundamental mode at 3.6 and its corresponding two harmonics. Excitation of similar SW modes triggered by OOP magnetic field excitation were also reported by Wang \textit{et al.}~\cite{wang2020effect}.
\vspace{-1em}\begin{table}[h]
    \centering
\caption{Exited SW modes in distinct stabilized chiral magnetic states for nanomagnet's diameter: 180\,nm.}
\label{tab:my_label}
    \begin{tabular}{|c|c|} \hline 
       Chiral states  & Frequency of excited SW modes (GHz) \\ \hline 
       RV  & 3.6, 6.6, 10.5 \\ \hline
       HS & 6, 6.7, 7.1, 7.7, 9.1\\ \hline  
       E-SKY  & 7.4 \\ \hline 
       SKYm  & 2.3, 12.5, 16.8 \\ \hline 
       E-SKYm  & 2.5, 8.7, 9, 16.1\\ \hline 
        Hopfion-like & 0.8, 2.7, 7.6\\ \hline
    \end{tabular}
\end{table}
As the system undergoes a transition to the precursor HS state, the rotational symmetry as observed in the RV state is disrupted. 
This results in the creation of multiple low-energy potential wells, possibly accounting for increased no. of SW modes i.e., four (see Table-I), with the strongest mode at the 7.1\,GHz.

Notably, frequency of the fundamental SW mode in the HS state at 170\,mT is greater than that of the RV state at 290\,mT ($\rm{f_{HS}>f_{RV}}$). This suggests that the effective field distribution in the nanomagnet is primarily influenced by intrinsic competing magnetic interactions, leading to distinct local spin configurations rather than the Zeeman field. Thus, the origin of the SW modes in the HS structure may differ and the specific local effective field of the HS state may introduce new SW modes of distinct character. Interestingly, as the system recovers the rotational symmetry  through the transitions to the E-SKY, SKYm, E-SKYm, and hopfion-like states, respectively, we observe characteristically different modes corresponding to all the spin textures. For example,a single fundamental mode at 7.4\,GHz (for $B_{\rm{ext, z}} = 130$\,mT) is excited for the E-SKY state, consistent with the earlier observations~\cite{song2019field}. 
For the SKYm state, three SW modes are excited. The weakest mode at 16.8\,GHz is shown separately in the inset of Fig.\,\ref{fig:5}(i). In general, in such $k\pi$ skyrmion states, no. of SW modes increases as a function of $k$~\cite{song2019field}. Thus, as the E-SKYm state is stabilized next with the external field further increased to the range of -138 to -158\,mT in the negative field direction, an additional N\'{e}el-type domain wall at the edges leads to the excitation of  total of four modes. In the case of the Hopfion-like state, three SW modes are excited with fundamental mode at 7.6\,GHz. The spectral charcterstics for hopfion-like state do not match with the above discussed spectra where SW modes at lowest frequency is typically appear as the fundamental mode except for E-SKym case. As noted above, the intensity or power of the SW modes for individual chiral magnetic state is related to the local effective field and therefore, except for RV state no systematic variation of mode-intensity as a function of either field or frequency is observed (see supplementary materials for details).  

Therefore, to discern the nature of the excited SW modes in the novel hopfion-like and its distinctiveness with respect to the other observed magnetic states, we have conducted an analysis involving the calculation of the spatial distribution of FFT power and phase of the excited SW modes within the discretized geometry. Figure\,\ref{fig:5}(ii) illustrates the spatially distributed FFT power (top) and phase (bottom) profiles of the SW modes shown in  Fig.\,\ref{fig:5}(i). 
This line power profile shown below each spatial power distribution elucidates the positional information of the radial nodes, labeled as `n'. On the other hand, the phase profile signifies the number of azimuthal nodes, labeled as `m'. Thus, the distinctive nature of a mode is characterized by its radial and azimuthal properties, quantified numerically through its radial and azimuthal nodes (n, m). In this analysis, it is pertinent to note that the nodal point at the center and the nodal ring at the periphery are omitted from consideration, as they arise from the Bloch point at the center and the geometrical confinement, respectively.

In general, OOP excitation in such chiral magnetic states gives rise to radially quantized and breathing modes, with FFT power typically manifesting as concentric circular rings separated by nodal rings~\cite{mruczkiewicz2017spin, song2019field, kim2014breathing, gareeva2016magnetic, vigo2021spin, awad2010precise}. The power profiles of all the SW modes excited for RV state exhibit this typical mode power distribution in concentric circular rings within the nanomagnet (as depicted in Figure\,5(ii.a)). For the fundamental SW mode at 3.6\,GHz, the mode power is prominently localized within the central region, with non-zero power everywhere else (n = 0) except at the center. It is important to note that at the core (center), the spins are oriented along the OOP direction, i.e., the +z direction (also the direction of perturbation), resulting in zero IP magnetization component. Consequently, the magnetic field perturbation does not induce any torque ($\tau\propto\theta$) at the core, leading to zero SW excitation at the center for magnetic states with an OOP polarized core.
Further, the phase profile of the corresponding mode confirms nearly coherent mode propagation (in-phase, m=0) throughout the geometry. Hence, the fundamental SW mode for the RV state can be identified as the radial SW mode (n=0, m=0). Higher-frequency modes observed at 6.6 and 10.5\,GHz are identified as higher-order radially quantized modes of (n=1, m=0) and (n=2, m=0) order, respectively. It is notable that considerable power is distributed along the edges of the nanomagnet for these modes. They also exhibit a phase jump of approximately 180$^\circ$ across nodal lines, a phenomenon absent in the fundamental SW mode at 3.6\,GHz. The significant power at the nanomagnet's edge arises due to the presence of tilted spins resulting from non-trivial boundary conditions imposed by the DM interaction. Similar characterization and analysis of the SW modes excited for other chiral magnetic states have been performed and briefly discussed below. The modes excited in chiral states with similar characteristics are tabulated in table-\ref{tab:table2} for brevity.
\begin{table}[h]
    \centering
    \caption{Characteristics of SW modes excited in different chiral magnetic states as observed from the power and phase distribution shown in Fig.\,\ref{fig:5}(ii). Here `Y' and `N' denote `Yes' and `No', respectively.}
    \begin{tabular}{|m{6em}|m{6em}|m{6em}|m{7em}|}\hline
        Chiral state & no. of radial nodes & phase jump of $180^\circ$ & hybridization with azimuthal modes\\ \hline
        RV & 0, 1, 2 & N, Y, Y & N, N, N \\ \hline
        E-SKY & 1 & N & Y\\ \hline
        SKYm & 2, 2, 3 &N, N, Y & Y, Y, N \\ \hline
        E-SKYm & 3, 2, 3, 2 & Y, N, N, N & N, Y, Y, Y \\ \hline
    \end{tabular}
    \label{tab:table2}
\end{table}\\
In the E-SKY state, the fundamental SW mode at 7.4\,GHz is characterized as (n = 1, m=0) mode, similar to the 6.6\,GHz mode of the RV state. However, the SW mode power is distributed in a square-like manner within the nanomagnet, in contrast to the circular ring pattern. Similar squareness in the power profile is reported in the literature due to the possible hybridization between radial and azimuthal modes and due to the discretized geometry used in micro-magnetic simulations in such confined nanomagnets~\cite{booth2019collective, wang2020effect}. Thus, this deviation may be attributed to the hybrid nature of the mode, which can be further corroborated by the observed phase dependence on the azimuthal angle at the nodal ring (see Fig.\,\ref{fig:5}(ii.b)). The SW mode remains in phase except at the nodal lines, consequently lacking a \(180^\circ\) phase jump across the nodal ring.
Thus, SW mode appears to exhibit a hybridized between breathing and azimuthal mode. Such phase dependence on the azimuthal angle at the nodal lines has not been distinctly reported in previous research to the best of our knowledge. This sort of hybridization is also observed for the SW modes excited in SKYm and E-SKYm (refer table-\ref{tab:table2}). Additionally, we observe an inhomogeneous power distribution within the circular rings for the SW modes displaying phase dependence on the azimthal modes (see Fig.\,\ref{fig:5}(ii.b-d)).
\begin{figure*}
\includegraphics[width = 0.99\textwidth]{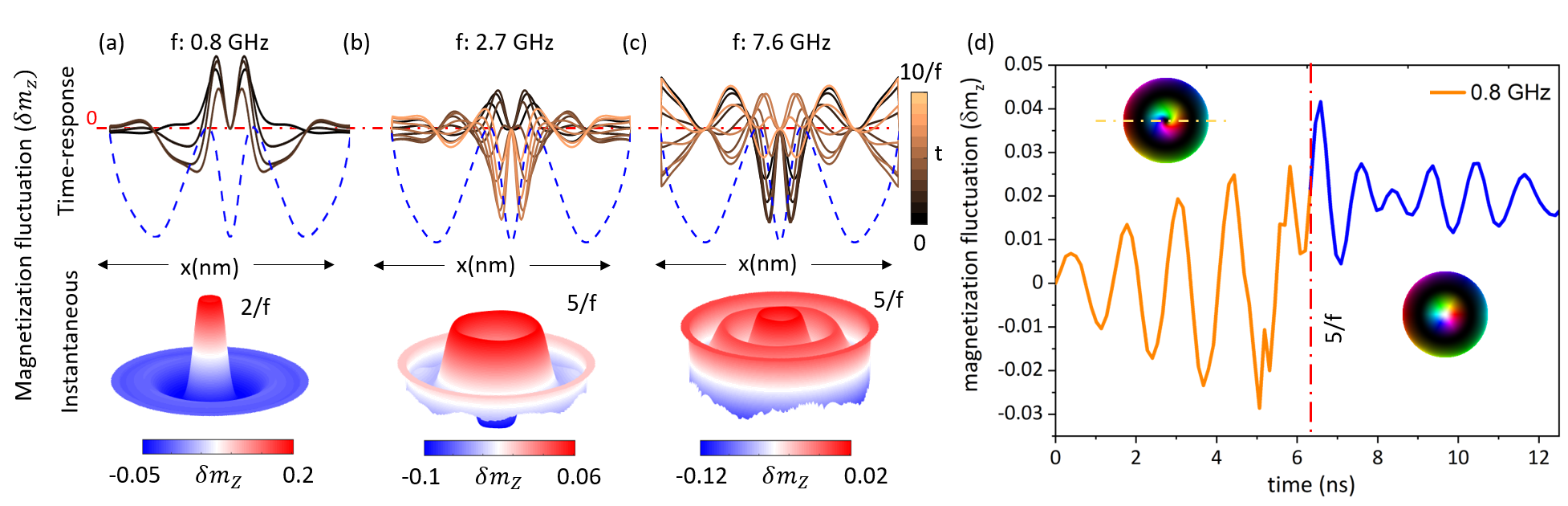}
\caption{\label{fig:6} (a-c) (top) The line profile illustrates the magnetization fluctuation in the $\delta m_z$ component across the geometrical center (radial direction) resulting from sinusoidal magnetic field wave excitation at SW resonance frequencies (f) of a hopfion-like state (-275\,mT). The profile is shown at distinct time instances with a time stepping of 1/f. The blue dotted line represents the z-component of magnetization in the equilibrium state (t = 0). (bottom) The image (surf plot) below each line profile displays the instantaneous magnetization fluctuation in the Co/Pt nanomagnet. (d) The average magnetization fluctuation ($\delta m_z$) resulting from a sinusoidal magnetic field wave at a frequency of 0.8 GHz, applied along the z-axis, indicates the transition from a hopfion-like to a skyrmion state. The inset depicts the magnetic configuration before and after the transition caused by the sinusoidal field application. The copper colorbar indicates the time parameter, while the BWR (blue-white-red) colorbar represents the magnitude of the magnetization fluctuation.\textit{Zero axis for both $M_z$ and $\delta m_z$}}
\end{figure*}
Typically, the nature of these SW modes i.e., breathing or radially quantized can be understood by examining the peak power distribution in the domain wall region of the chiral magnetic state. In the case of the breathing mode, the peak power of the SW mode is centered at the domain wall region (\(m_\mathrm{z}=0\)), whereas in the case of the radially quantized mode, the peak power is concentrated on the left and right sides of the domain wall (\(m_\mathrm{z}=0\)) with a node at the \(m_\mathrm{z}=0\) position.\par
Finally, we discuss in details the power and phase profiles of the three excited modes for the novel Hopfion-like chiral state.
We observe two important characteristics. First, the core region identified as the region with no (or very weak) excitation, is relatively small for this state.  Secondly, as shown in  Fig.\,\ref{fig:6}(ii.e), the SW power is distributed in the form of concentric circular rings with significant power localized near the core. Careful analysis of the power distribution of these modes show regions of gradually decreasing power immediately beyond the brightest ring. This appears as a shoulder next to the peak in the line profile and is clearly visible for the 0.8\,GHz mode (see LP8 in Fig.\,\ref{fig:5}). Due to the appearance of additional nodes for 2.7 and 7.6\,GHz modes, this shoulder region is relatively small. Such shoulders are exclusively observed for the Hopfion-like texture. Our analysis do not suggest any such power profile for other chiral states.
Concurrent examination of the underlying spin configuration suggest that this gradual decrease in power begins at the position of Bloch-type spin rotation in domain wall region.
Further, the line profiles clearly show that these modes exhibit radial nodes nos., n= 1, 2, and 2. 

The phase profile of this novel state shows a markedly different behavior than observed for the other textures. Interestingly in this case, we observe a continuous variation of SW phase along the radial direction persisting up to the first nodal ring position suggesting counter-propagating SWs of different wavelengths at the same frequency. This phenomenon is known as non-reciprocity in SWs and was earlier observed for domain walls or spin configurations with spins not parallel to wavevector, stabilized in the presence of DM interaction~\cite{garcia2015narrow}. Thus, the non-reciprocity in this case can be attributed to the varying helicity of the spins in the Hopfion-like state that we observe here. Furthermore, similar to the cases of RV, SKYm and ESKYm states, the phase profiles for the 0.8 and 2.7\,GHz mode for this Hopfion-like state also exhibit phase jumps of \(\sim 180^\circ\) across the nodal ring. However, SW mode at 7.6\,GHz does not exhibit any such phase jump. In order to determine the nature i.e., breathing or radially quantized, of the three observed modes, we carry out further analysis of time evolution of oscillations in z-component of magnetization ($\delta m_z(t) = m_z(t) - m_Z(0)$). 

\begin{figure}
\includegraphics[width = \linewidth]{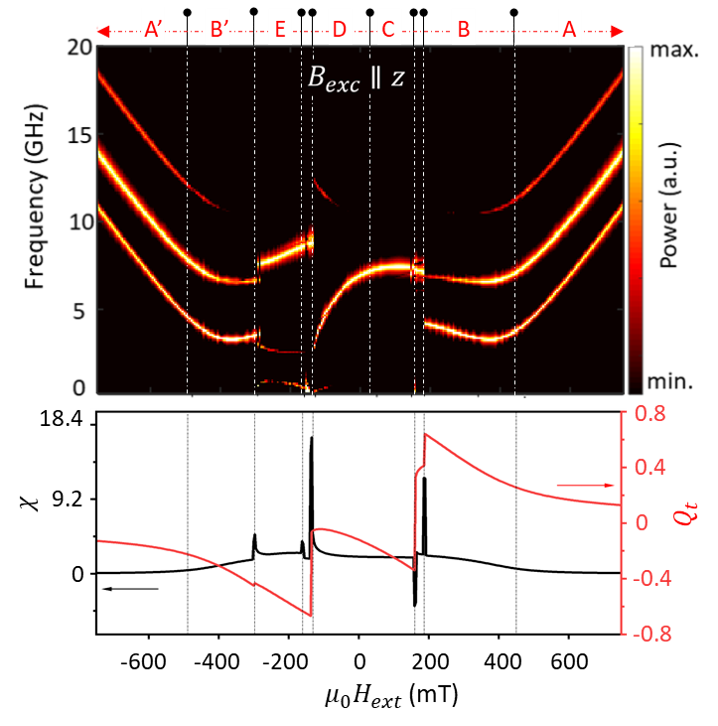}
\caption{\label{fig:7} The resonance frequencies plotted during the transitions to RV,
HS, E-SKY, SKYm, E-SKYm, hopfion-like, and RV (opposite polarity) states as $\mu_0H_{\rm{ext}}$ is downswept. The static properties including topological charge and susceptibility of these states are presented below.}
\end{figure}

For each SW mode, the evolution of $\delta m_z$ is calculated by resonating the equilibrium state by applying a sinusoidal magnetic field at the corresponding mode frequency, applied along the OOP direction. Fig.\,\ref{fig:6}(a-c) show the evolution of $\delta m_z$. Different colors in the line plots represent different time instances (1/10f) at which magnetization oscillations are averaged over ten consecutive wavelengths of the sinusoidal magnetic field. For clarity, instantaneous 3D magnetization oscillation profiles calculated at a particular instance are shown below each line plot.
The profiles of magnetization oscillation for these modes show that the oscillations $\delta m_{z}$ lack a distinct node or extremum at the domain wall position (\(m_z = 0\)) which is apparent for 0.8\,GHz mode. Additionally, the magnetization oscillation profile across the radial direction (from the center to the first node) vary gradually at different time instances. This observation suggests that the standing SW modes do not remain still due to the non-reciprocity of SW. Further, we observe a consistent node (zero oscillation) localized in the OOP-magnetized region (\(m_z = -1\)) in all the three SW modes. In case of 2.7 and 7.6\, GHz, we observe that the additional node is present at $m_z\neq 0, 1$ position and is localized closed to domain wall region. The presence of this additional node at this position indicate the geometrical quantized nature of the SW modes.
For clarity, a movie animation illustrating the magnetization oscialltion within the nanomagnet is provided in the supplementary information. Interstingly, the analysis based on the excitation of resonant SW modes through sinusoidal field pulses reveals a noteworthy phenomenon of magnetic state transformation via core polarity switching. This transformation is illustrated in Fig.\,\ref{fig:6}(d), where the magnetization oscillation for the hopfion-like state is presented when a sinusoidal magnetic field at frequency `0.8\,GHz' is applied in the OOP direction. The underlying hopfion-like state undergoes a core reversal during excitation, transitioning from a negative to a positive z-direction. Additionally, the hybrid domain wall rotation transforms into a pure-N\'{e}el type rotation, resulting in the emergence of an SKY state. This transition becomes evident after five consecutive sinusoidal magnetic field oscillations (t = 5/f). Interestingly, during the magnetic field sweep discussed above, the hopfion-like state is transitioned to a RV state in contrast to SKY state. Indeed, the SKY state is not observed during the complete field sweep (see Fig.\,\ref{fig:3}). Thus, our study demonstrates an energy-efficient approach to manipulate the underlying chiral magnetic state, a highly sought-after characteristic in the realm of spintronic memory devices.
So far, the behavior of SW modes for distinct chiral magnetic states are discussed at a particular magnetic field strength. However, these magnetic states are transformed from one state to another gradually and abruptly during magnetic field sweep (see phase diagram in Fig.\,\ref{fig:3}). Thus, a complete understanding of the evolution of the mode's frequency as the underlying magnetic state transformed into another state can not be thoroughly understood from the above analysis. Therefore, for the comprehensive study. we have performed detailed calculations of SW spectra during magnetic field sweep from 1 to -1\,T at a field step of 5\,mT (see Fig.\,\ref{fig:7}). The calculated dispersion of the SW mode' frequency w.r.t. $\mu_0H_{\rm{ext}}$ is found to be asymmetric across the remanence. This asymmetry is attributed to the variable magnetic states stabilized within this magnetic field range. Additionally, dispersion plot demonstrates regions of distinct $df/\mu_0dH_{\rm{ext, z}}$, which are segregated by white dashed lines in the plot. These regions correspond to distinct magnetic states observed as an energy-minimized ground state while sweeping the magnetic field. From the visual inspection of the frequency response as a function of the $\mu_0H_{\rm{ext}}$, we note the following interesting observations: 1) In regions A and A$^\prime$, three SW modes are observed which vary almost linearly with the $\mu_0H_{\rm{ext}}$, viz., $f\propto \mu_0H_{\rm{ext, z}}$. The linear relationship between the mode frequency and the magnetic field strength infers the dominant contribution of the Zeeman field in the total effective magnetic field. Thus, the observed SW modes may therefore be conceived as Kittle modes, i.e., SW modes governed by the following expression in a confined nanomagnet:
\begin{equation}
    f = \gamma\sqrt{(\vec{B}_{\rm{eff}}+(N_{\rm{x}}-N_{\rm{z}})M_{\rm{s}})(\vec{B}_{\rm{eff}}+(N_{\rm{y}}-N_{\rm{z}})M_{\rm{s}})}
\end{equation}
These modes were characterized as radially quantized modes from the power and phase profiles in the above discussions. 2) In regions B and B$^\prime$, stabilization of the radial vortex state results in a saddle point in the frequency response curve. Further, the frequency of the SW modes varies nearly linearly with $\mu_0H_{\rm{ext}}$ in region B, exhibiting a negative slope. The observed non-trivial response of the SW modes suggests a change in the underlying energetics associated with chiral magnetic states. This change can be attributed to the competing magnetic interactions in addition to relatively weak Zeeman interaction, which started playing a vital role in determining the minimum-energy ground state. It may be understood that the specific arrangement of spins in chiral magnetic state, is giving rise to an internal field that is majorly conditioning the entire effective field distribution within the nanomagnet. Consequently, the effect of the Zeeman field on the total effective field appears to be annulled by the contemporary effect of the emergent internal field from the stabilized chiral magnetic state.  3) From regions B to D, the response of the mode' frequency has altered significantly as the underlying magnetic state is transfigured in to E-SKY from RV state through a precursor HS state (see Fig.\,\ref{fig:1}). A narrow window between regions B and C become apparent where a metastable HS state is stabilized. In this narrow window, a fundamental SW mode at 7.1\,GHz along with a few weak SW modes at nearby frequencies are apparent. As discussed above, these weak SW modes may appear due to broken rotational symmetry of the spins configuration in the HS state. As the system reinstates the rotational symmetry via the formation of the E-SKY, only fundamental SW mode prevails at $\sim$7.4\,GHz through out the region-iii of Fig.\,\ref{fig:1}(a), which demonstrates minor variation in its frequency within its stabilization region. This further corroborate that the emergent internal magnetic field response is peculiar to the underlying chiral magnetic state. 4) Interestingly, salient features as new modes and distinct f-$\mu_0H_{\rm{ext}}$ reponse appear inregion D, which account for the E-SKY to SKYm transition. Notably, we observe a seemless transition of E-SKY to SKYm in $M, \chi$, and$Q_t$ vs. $\mu_0H_{\rm{ext}}$ plot (see Fig.\,\ref{fig:1} or bottom pannel of fig.\,\ref{fig:7}). 
5) In region E, where the novel hopfion-like state is stabilized, all three SW modes discussed above persist within its stability region, however with distinct frequency response. The response of the fundamental mode (strong mode, shown in bright color in Fig.\,\ref{fig:7}) is similar to the SW modes excited in the RV state (region B). The power and phase profiles discussed above suggest that this is indeed a new SW mode specific to the hopfion-like state. The other weak SW modes also display non-trivial relation between the mode frequency and the $\mu_0H_{\rm{ext}}$. The observations thus suggest that SW mode spectra can not be directly explained in terms of the externally applied Zeeman field. The emergence of new modes and their response against an external magnetic field is peculiar to the underlying chiral magnetic state. Consequentially, a clear picture of the phase transitions of the underlying magnetization states can be procured from the frequency response of SW modes as a function of $\mu_0H_{\rm{ext}}$.\par
In summary, our extensive micro-magnetic simulations conducted on ultra-thin Co/Pt nanomagnets have provided details regarding the rich phase diagram of different chiral magnetic states as a function of nanomagnet's diameter (d) and the applied external magnetic field. One noteworthy finding was the observation of a novel chiral magnetic state whose spin configuration closely resembles the IP and OOP component of magnetization distribution in 3D magnetic soliton viz., Hopfion~\cite{luk2020hopfions, kent2021creation}. Our analysis of the underlying energy landscape revealed that this Hopfion-like state is stabilized through the minimization of demagnetization energy. This finding underscores the critical role played by demagnetization energy in stabilizing chiral magnetic states, characterized by twisted or hybrid domain walls. Furthermore, our investigation reveals the presence of nonreciprocal SW mode behavior specific to the Hopfion-like state, a feature notably absent in other stabilized chiral magnetic states. Such nonreiprocity is exploited in the developement of SW based dioded and SW conduits. Additionally, we observe hybridization of azimuthal modes with characteristic breathing and radial quantized SW modes for all the stabilized chiral magnetic states. Our comprehensive analysis of the SW behavior of the underlying magnetic states during external field sweeps revealed distinctive patterns in SW behavior accompanying transitions in the magnetic state. This observation underscores the pivotal role of SW responses in providing clear insights into phase boundaries, even for transitions that might appear seamless in magnetization and topological charge responses. Importantly, we demonstrate a fascinating phenomenon wherein the Hopfion-like state is switched to a skyrmion state within nanoseconds via it's breathing SW mode at 0.8\,GHz. Such direct switching between the magnetic states has not observed previously. However, the core polarity switching is discussed in the literature. This intriguing phenomenon holds significant promise for the development of power-efficient memory devices and re-configurable magnonics where high degree of SW tunability is sought.

\begin{acknowledgments}
We acknowledge financial support from Science and Engineering Board Research Board, (Govt. of India) through grant nos. CRG/2018/004340 and SPR/2021/000762. Also, High-Performance Computation (HPC) facility of IIT Delhi is acknowledged for providing a platform to run the micromagnetic simulations. NA and YK thank the Ministry of Education, Government of India, for the research fellowship.
\end{acknowledgments}

\bibliography{main_text}

\end{document}